# The Effect of Pattern Quality on Measurements of Stress Heterogeneity and Geometrically Necessary Dislocation Density by High-Angular Resolution Electron Backscatter Diffraction

Harison S. Wiesman[1,2] and David Wallis[1]

[1]Department of Earth Sciences, University of Cambridge, Cambridge, UK
[2]Now at School of Earth and Environmental Sciences, University of Minnesota, Minneapolis, MN, USA

**Abstract**

We examine the effect of pattern quality on the output of high-angular resolution electron backscatter diffraction (HR-EBSD) analyses. Band contrast, as a proxy for pattern quality, was varied by adjusting the number of frames averaged per electron backscatter pattern during data collection. The same region in a deformed sample of the mineral olivine was mapped six times varying the number of frames averaged between 1 and 30 between each map. Each data set was analyzed with HR-EBSD, producing maps of intragranular stress heterogeneity and geometrically necessary dislocation (GND) density. As the number of frames averaged increased, the noise in stress and GND calculations decreased, revealing more substructure in the mapped region. The worst pixels, with low band contrast, are the most improved by increased frame averaging, whereas those with high band contrast are largely unaffected. Additionally, the probability distribution of stresses narrows as high-stress noise is reduced with increased pattern quality, which also affects estimates of dislocation density from statistical analysis of the stress distributions. As regions with high stress and/or high GND density are typically of interest in HR-EBSD maps and are often associated with low band contrast, frame averaging may be used as a tool to improve the quality of these analyses. Most importantly, however, is that comparisons are made between HR-EBSD datasets with similar mean band contrast to ensure that observed differences are microstructural in origin and not an artefact of data collection.

**Highlights**
- Increasing frame averaging reduces noise in HR-EBSD analyses.
- Pixels with low band contrast are the most improved by increasing frame averaging.
- Varying frame averaging changes the statistics of results.
- Comparisons among data sets should employ similar pattern quality.

# 1. Introduction

Electron backscatter diffraction (EBSD) is a widely used tool for characterizing engineering, geological, and biological materials (e.g., Prior et al., 1999; Griesshaber et al., 2010; Wilkinson and Britton, 2012). During characterization, electron backscatter patterns (EBSPs) are collected in a scanning electron microscope (SEM) and are used to determine the crystallographic orientation at each point. This mapping allows for microstructural characteristics, such as orientation gradients, grain size, and/or grain shape, to be determined and analyzed. However, the Hough transform-based analysis typically used by commercial EBSD systems to identify phases and orientations does not measure the distortion among EBSPs themselves, which contain valuable information about the elastic strain stored in the lattice (Wilkinson et al., 2006).

High-angular resolution electron backscatter diffraction (HR-EBSD) improves on this aspect of EBSD by making more sensitive and direct analyses of the EBSPs (Wilkinson et al., 2006; Britton and Wilkinson 2011, 2012). HR-EBSD is a post-processing, cross-correlation based technique that compares the positions of features within regions of interest in each pattern to those same regions in a reference pattern, yielding information about the relative distortion of the EBSPs across a crystal (Britton and Hickey, 2017). Shifts in the regions of interest due to these distortions are quantified to sub-pixel accuracy, allowing the precisely measured distortions to be related to the elastic strains and lattice rotations within a deformed material (Britton and Wilkinson, 2012; Wallis et al., 2019). HR-EBSD provides up to two orders of magnitude better angular resolution than that of Hough transform-based EBSD, reducing noise in estimates of geometrically necessary dislocation (GND) density (Wilkinson et al., 2006; Jiang et al., 2013).

Essential to the precise image analysis used with HR-EBSD is the use of high-quality EBSPs (Wilkinson and Randman, 2010, Britton et al., 2010). In addition to having high pixel resolution and bit depth, bands present in EBSPs should also be well defined; that is, band edges should be sharp and bands should have high contrast relative to the background noise. If the bands are not well defined, higher degrees of noise are introduced into the analysis as it may be difficult to accurately determine shifts between EBSPs. Band definition can be influenced by many factors including the quality of the surface polish, the inherent crystal chemistry and structure, and/or the composition and thickness of conductive coats required by many insulating materials. One way to improve the quality of EBSPs is to employ frame averaging during data collection. At the cost of collection time, this procedure averages multiple EBSPs collected at each point, reducing noise as more EBSPs are included and improving the definition of features within the resultant EBSP. To test the effect of pattern quality on analyses associated with HR-EBSD, we collected six HR-EBSD datasets from a single location in a sample of the mineral olivine (($Mg_{0.9}$, $Fe_{0.1}$)$_2$SiO$_4$) with different numbers of frames averaged. We then assessed features of the resultant maps to determine the impact of pattern quality on these analyses in a manner that can guide decisions during future data acquisition.

## 2. Methods

### 2.1. Sample Preparation

We mapped a deformed sample of San Carlos olivine for analysis with HR-EBSD. This sample was deformed in torsion in a high-resolution, gas-medium apparatus (Paterson and Olgaard, 2000) at a temperature of 1523 K, confining pressure of 300 MPa, and at a constant shear strain rate of $1.5 \times 10^{-4}$ s$^{-1}$. The final shear stress and shear strain reached in this experiment were 125 MPa and 0.41, respectively. After deformation, the sample was polished with diamond lapping film down to 0.5 μm grit, then finished with a chemo-mechanical polish using 40 nm colloidal silica. The sample was coated with 6 nm of carbon prior to data collection to prevent charge build-up in the SEM. Further details of the deformation experiment and surface preparation are given by Wiesman et al. (2024). This sample was selected for the present study due to its varied intragranular microstructure so that the effect of pattern quality on the appearance and magnitude of different features could be easily assessed.

### 2.2. Data Collection with EBSD

EBSD data were collected in a Zeiss Gemini 300 SEM at an accelerating voltage of 30 kV with an Oxford Instruments Symmetry detector using the AZtec 4.0 data acquisition software in the Wolfson Electron Microscopy Suite at the University of Cambridge. EBSD maps were collected with a step size of 0.25 μm over an area of $75 \times 75$ μm$^2$. During data collection, patterns were saved at a resolution of $1244 \times 1024$ pixels for analysis with HR-EBSD. Both static and automatic background corrections were applied to EPSPs by the software as they were collected. The microscope was calibrated for HR-EBSD following Wilkinson et al. (2006) and reference-frame conventions were validated following Britton et al. (2016). To exclusively vary the pattern quality, the six maps were from the same sample area, each with a different number of frames averaged. Specifically, maps were made with 1, 3, 5, 10, 15, and 30 frames averaged. Due to the required acquisition time, the map collected with 30 frames averaged had to be collected in a different session, such that the area mapped largely overlaps with, but is not exactly the same as the area featured in the other maps. In addition, sample or beam drift in the remaining five maps may have caused slight shifts on the order of a few measurement points between each map. Olivine is stable under the electron beam in the SEM and therefore we assume that any effect of beam damage on the quality of the diffraction patterns is negligible.

To aid in our analysis, each map was assigned a mean band-contrast (BC) value, also sometimes referred to as an image quality factor (Wright and Nowell, 2006 and references therein). At each point in a given map the BC was calculated from the peak height of bands in Hough space relative to the background. The mean BC was then calculated as an average across all points in a given map. Because these values are automatically calculated by the AZtec software, the mean BC is a useful first indicator of pattern quality as it is reported during data collection.

## 2.3. Analysis with HR-EBSD

Saved EBSPs were analyzed using HR-EBSD. HR-EBSD is a cross-correlation technique that compares the positions of regions of interest (ROIs) within each pattern to those in a reference pattern. Further details of the technique are provided by Wilkinson et al. (2006) and Britton and Wilkinson (2011, 2012) and applications to geological materials are discussed by Wallis et al. (2019). For analysis of each map, reference points were manually selected and 100 ROIs, each 256 × 256 pixels in size, were automatically selected in each diffraction pattern. ROIs were then filtered in Fourier space to reduce noise before the cross-correlation (Wilkinson et al., 2006). Additionally, the data were filtered to remove poor quality points. We filtered out results for which the normalised peak height in the cross-correlation function was < 0.3 and those with a mean angular error in the deformation gradient tensor > 0.004 (Britton and Wilkinson, 2011). To facilitate comparisons between maps with different amounts of frame averaging, we used the same reference-pattern locations in each map. Finally, because of the relatively large bulk strain experienced by this sample, we followed the pattern remapping procedure described by Britton and Wilkinson (2012), allowing for greater accuracy of the elastic strain measurements for points with lattice rotations greater than 1˚.

This cross-correlation procedure maps shifts in the positions of the ROIs relative to their positions in the corresponding reference pattern. A deformation-gradient tensor is then fit to these shifts. The symmetric and antisymmetric components of this tensor correspond to the elastic strains and lattice rotations, respectively (Wilkinson et al., 2006; Britton and Wilkinson, 2011, 2012). These elastic strains were converted into stresses using Hooke's law and the elastic properties of olivine (Abramson et al., 1997). The resultant stresses are calculated relative to the unknown stress state at the reference point, so, to simplify the interpretation of the stresses, we normalized them by subtracting the mean of each component of the stress tensor within each grain from the corresponding component at each point within that grain. In this way, the reported stresses represent the stress heterogeneity relative to the mean stress state of each grain (Jiang et al., 2013; Mikami et al., 2015; Wallis et al., 2017, 2019). In the present study, we focus only on the $\sigma_{12}$ component of the stress tensor as this component is the least modified by sectioning the sample and is of particular relevance to deformation processes (Wallis et al., 2019). Similarly, the lattice rotations were converted into densities of GNDs by fitting densities of the six dominant dislocation types in olivine to the measured lattice curvature (Wallis et al., 2016).

## 2.4. The Restricted Second Moment and Average Dislocation Density

To further analyze the stress fields generated from HR-EBSD analyses, recent studies have examined the restricted second moment of the stress distributions (e.g., Wilkinson et al., 2014; Wallis et al., 2021). Similar to peak broadening in X-ray diffraction, if stress heterogeneity present in the sample is due to stress fields around dislocations, the probability distribution, $P(\sigma)$, of stress heterogeneity in a map should carry information about the dislocation content (Groma 1998; Wilkinson et al., 2014; Kalackska et al., 2017). Specifically, Wilkinson et al. (2014) suggested that the tails of the stress distribution should decay as $P(\sigma) \propto |\sigma|^{-3}$. If this relationship is correct, the restricted second moment, calculated as $\nu_2 = \int_{-\sigma}^{+\sigma} P(\sigma')\sigma'^2 d\sigma'$, should be proportional to ln($\sigma$) along the distribution tails. Assuming that the dislocations are straight edge dislocations and are non-interacting, i.e. each dislocation controls the stress field in its immediate vicinity, the constant of proportionality, *m*, coupling $\nu_2$ to ln($\sigma$) should be related to the total dislocation density as

$$m = \frac{(Gb)^2}{8\pi(1-v)^2}\rho \quad (1),$$

in which *G* is the shear modulus, *b* is the Burgers vector, *v* is Poisson's ratio, and *ρ* is dislocation density. For olivine, *G* = 70 GPa , *b* = 5 × 10$^{-10}$ m, and *v* = 0.25 (Deer et al., 1992, pp 5; Zhao et al., 2018). This analysis provides a method to estimate the total dislocation density and to tie the source of stress heterogeneity to dislocation stress fields directly (Kalacska et al. 2017; Wallis et al. 2021). We note, however, that the arrangement of, and elastic interactions between dislocations are likely more complicated in reality than assumed by this simple model. Nevertheless, our focus here is simply on quantifying the extent to which pattern quality affects $\nu_2$ and *m*.

## 3. Results
### 3.1. EBSD Data

Band contrast maps are displayed in Figure 1 along with a representative EBSP from the same location in each map. In the maps in Figure 1a, each pixel is shaded by its BC. Variations of the BC within grains are related to distortion of the crystal lattice due to GNDs. The values of mean BC for each map are reported in Table 1 and increase from 57 to 198 as the number of frames averaged increases from 1 to 30. Qualitatively, in Figure 1b, the definition of features in the EBSPs increases from 1 frame to 30 frames averaged, further demonstrating that the pattern quality has increased.

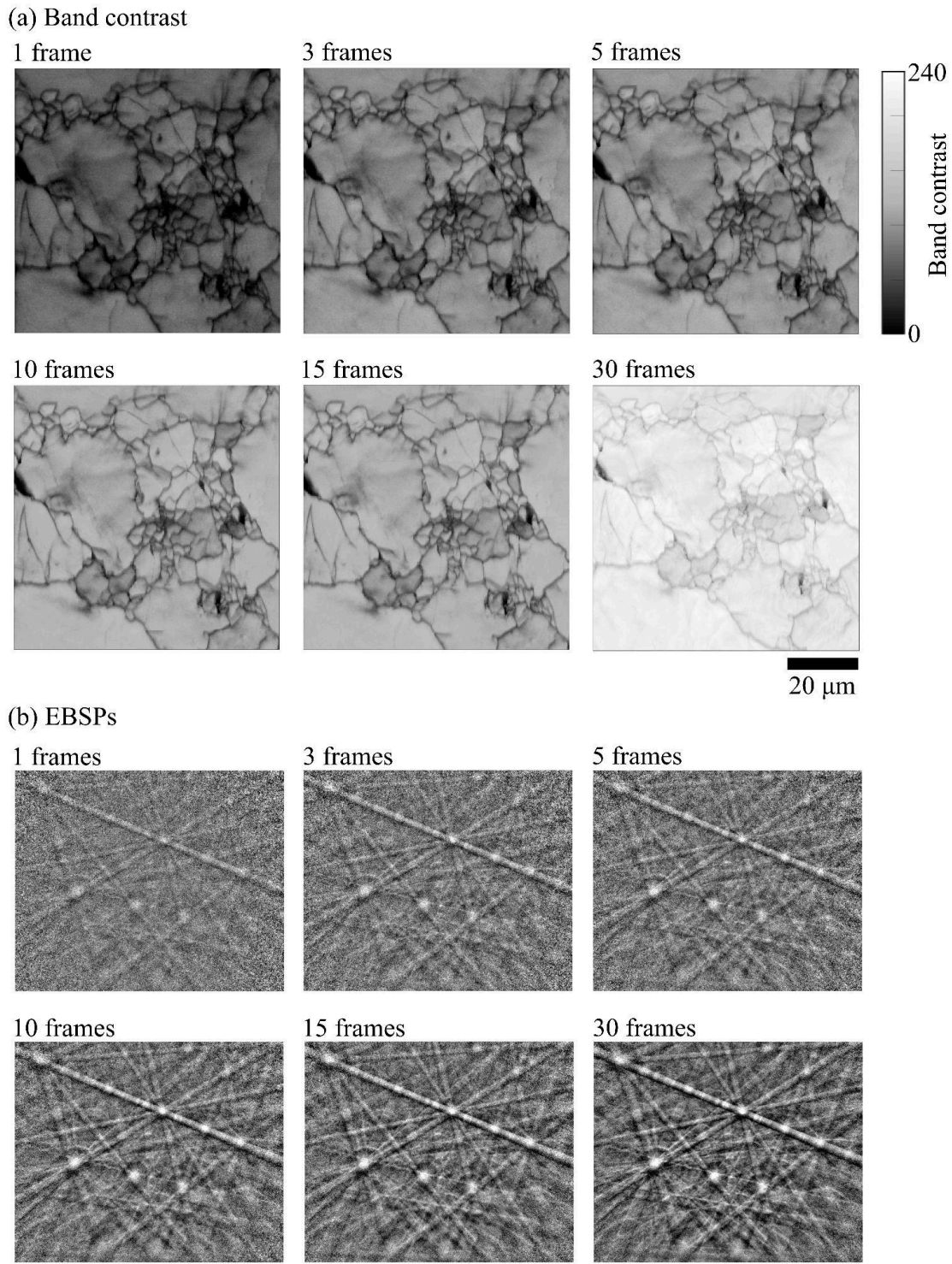

Figure 1. (a) Band contrast and (b) EBSPs from each map collected with different numbers of frames averaged per pattern.

| Number of frames averaged | Mean band contrast | $m$ (GPa$^2$) | $\rho$ (×10$^{15}$ m$^{-2}$) | Signal-to-noise ratio of bad pattern (×10$^3$) | Signal-to-noise ratio of good pattern (×10$^3$) |
| --- | --- | --- | --- | --- | --- |
| 1 | 57 | 0.17 | 6.1 | 0.09 | 0.08 |
| 3 | 84 | 0.12 | 4.3 | 0.22 | 0.18 |
| 5 | 91 | 0.11 | 4.1 | 0.25 | 0.22 |
| 10 | 107 | $9.9 \times 10^{-2}$ | 3.5 | 0.63 | 0.50 |
| 15 | 112 | $9.9 \times 10^{-2}$ | 3.5 | 0.95 | 0.73 |
| 30 | 198 | $8.9 \times 10^{-2}$ | 3.2 | 0.99 | 1.3 |

Table 1. Summary of quality metrics and the restricted second moment derived from each map.

### 3.2. HR-EBSD Data

Figure 2 presents maps of intragranular stress heterogeneity and GND density from the HR-EBSD analyses. In the maps of intragranular stress heterogeneity in Figure 2a, regions of high stress magnitude reaching up to ±1 GPa are present as patches and linear features within grains. These features are visible in each of the maps regardless of the number of frames averaged. However, the map with only one frame averaged displays a significant amount of noise, manifesting as large differences in stress between neighboring points. This noise obscures many of the more subtle features within grains that are visible in maps with more frames averaged. Qualitatively, the magnitude of this noise decreases as the number of frames averaged increases. No major qualitative differences are present in the degree of noise amongst maps with five or more frames averaged, although features within the stress maps continue to change in intensity.

The maps of GND density in Figure 2b display similar trends to the maps of stress heterogeneity. Patches of elevated GND density are present within grains, typically near grain boundaries, reaching up to $10^{14}$ m$^{-2}$. Linear features corresponding to subgrain boundaries typically reach GND densities of $\geq 10^{15}$ m$^{-2}$. Some grains in each map have greater GND densities than others, potentially due to particular grain orientations generating higher noise floors (Wallis et al., 2016, 2019). As with the maps of stress heterogeneity, noise in the GND map with only one frame averaged obscures some of the more subtle features observed in the other maps. The noise levels decrease as more frames are averaged per pattern, although the most features are clearly visible in the maps with 10 or more frames averaged.

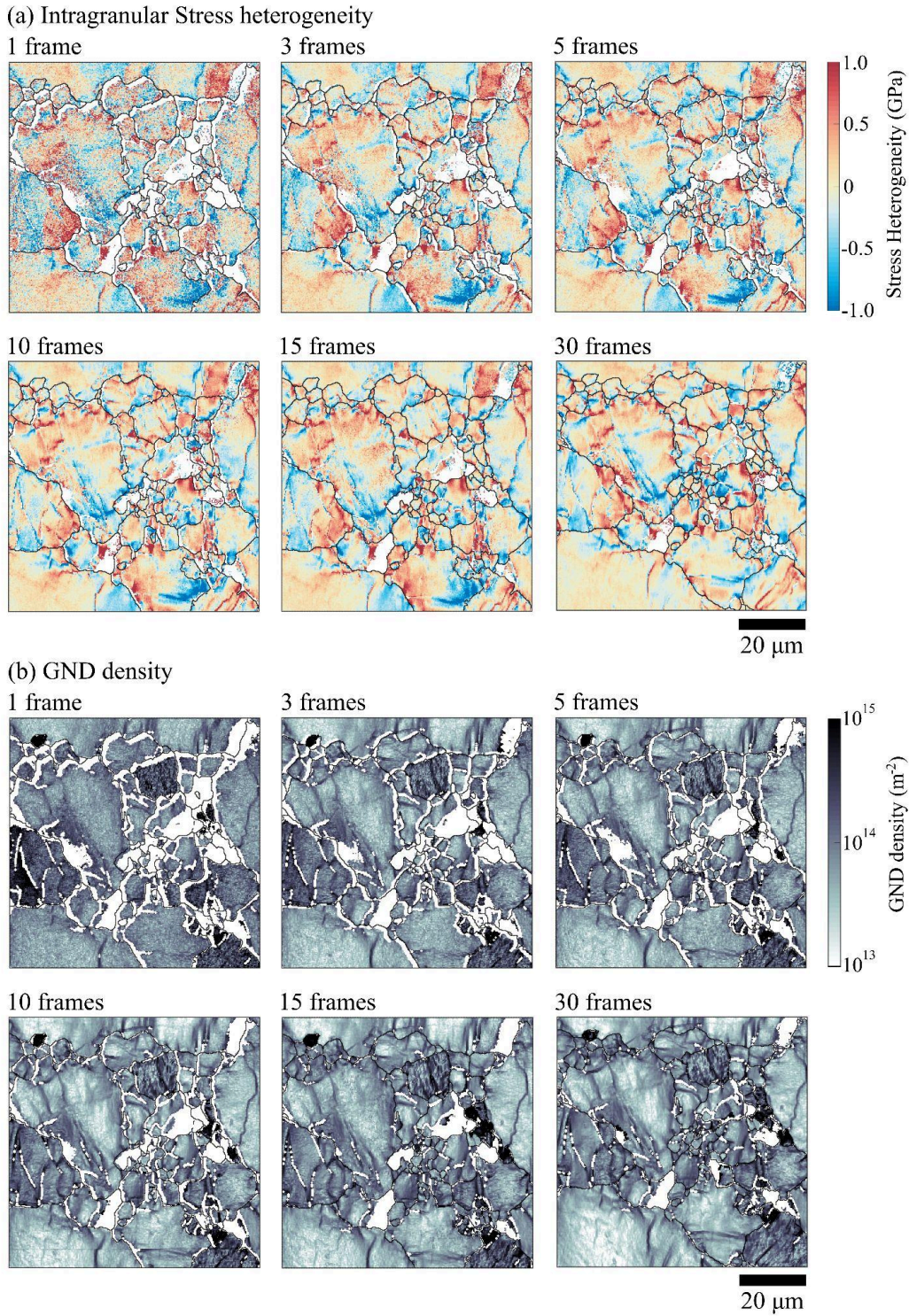

Figure 2. Maps of (a) intragranular stress heterogeneity and (b) GND density from HR-EBSD. Grain boundaries are plotted as black lines. White points in each map either did not index during data collection or failed the quality criteria during HR-EBSD analysis.

### 3.3. Stress Distributions and their Restricted Second Moments

Probability distributions from the stress maps in Figure 2a are plotted in Figure 3 alongside normal probability plots and the restricted second moment calculated from each of the distributions. The distribution and restricted second moment for an undeformed Si standard (Wallis et al., 2022) are also plotted in red for comparison. As the number of frames averaged increases, the probability distribution of stresses in Figure 3a narrows, indicating a smaller proportion of high-stress noise in the maps created from EBSPs with more frames averaged. The narrowing of these distributions is also quantified in Figure 3b by measuring the 95$^{th}$ percentile of each distribution, which decreases from 960 MPa to 640 MPa as the number of frames increases. In turn, the change in shape of the distribution results in a steeper normally distributed, linear portion of the normal probability plots in Figure 3c. The cumulative probability distribution for Si is entirely linear, indicating that the map was composed of normally distributed noise. In contrast, the high-stress tails of the cumulative probability distributions for the olivine maps deviate from the linear shape expected for a normal distribution. The deviation from a normal distribution indicates that the greatest stresses can instead be described by a different distribution. We tested for the presence of a $P(\sigma) \propto |\sigma|^{-3}$ relationship by calculating $v_2$ from the probability distributions, which is plotted against $\ln(\sigma)$ in Figure 3c. Each distribution has a linear segment of $v_2$ versus $\ln(\sigma)$ that extends between $\ln(\sigma / [1\ \text{GPa}]) = 0.2$ and 1. The slope of this linear segment, $m$, is reported in Table 1 along with the dislocation densities, $\rho$, calculated from this slope using Equation 1. The proportionality between $v_2$ and $\ln(\sigma)$ over this range of intermediate stresses suggests that part of the stress heterogeneity observed in Figure 2a is caused by the stress fields of dislocations. Deviations from this linear relationship at the greatest stresses are potentially caused by spatial averaging of very localized, high-magnitude stress in the interaction volume of the electron beam (Kaláckska et al., 2017), while nonlinearity at low stresses is likely related to the normally distributed portion of the stress distribution (Wilkinson et al., 2014, Wallis et al., 2021). As the number of frames averaged per pattern increases from 1 to 30, the value of $m$ decreases from 0.17 GPa$^2$ to 8.9 × 10$^{-2}$ GPa$^2$. Using Equation 1, this decrease in $m$ leads to a corresponding decrease in apparent dislocation density by approximately a factor of two from 6.1 × 10$^{15}$ m$^{-2}$ to 3.2 × 10$^{15}$ m$^{-2}$.

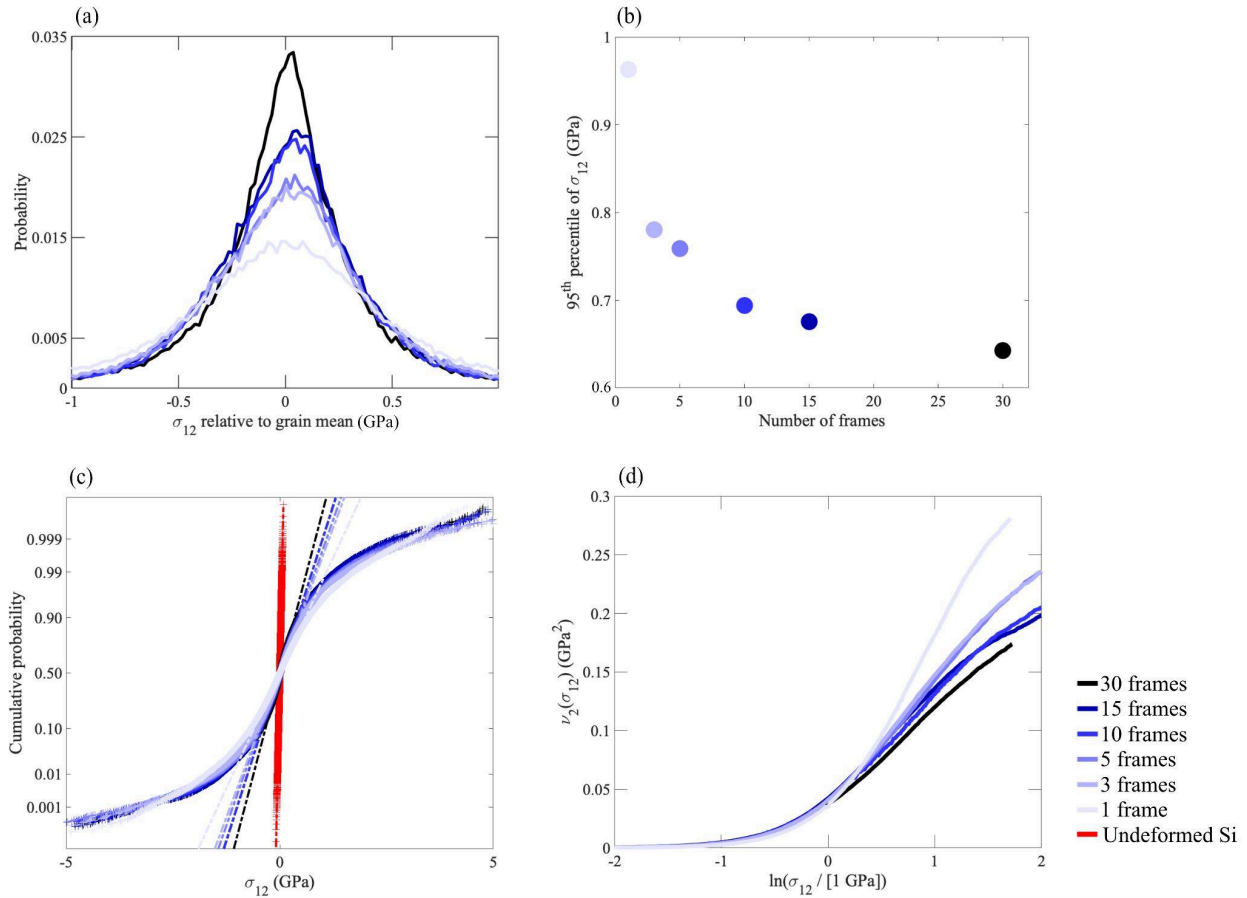

Figure 3. (a) Probability distributions of stress heterogeneity in Figure 2. (b) 95$^{th}$ percentile of the probability distributions in (a) versus the number of frames averaged per point in a given map. (c) Normal probability plots of the stress distributions in (a). The cumulative-probability axis is scaled such that a normal distribution is a straight line. The red curve represents data from a piece of undeformed Si for comparison (Wallis et al., 2022). (d) Restricted second moment versus ln($\sigma$) for each of the stress distributions in (a). Colors from light blue to black represent results from maps with different numbers of frames averaged.

## 4. Discussion
### 4.1. Assessing the Effect of Pattern Quality on HR-EBSD Data

From the results in Figures 2 and 3, it is clear that pattern quality, and hence frame averaging, has an effect on the results of HR-EBSD analysis and their interpretation. In the following sections, we characterize the effect of frame averaging in a way that may serve practical decision making for future data acquisition. We employ two methods to quantify the difference that increased pattern quality makes to analysis with HR-EBSD: examining the signal-to-noise ratio of radial power spectra calculated directly from the EBSPs and then by comparing differences between the same points in HR-EBSD maps of both stress heterogeneity and GND density with different amounts of frame averaging.

### 4.1.1 Radial Power Spectrum and Signal-to-Noise Ratio

To better quantify the change in quality of the EBSPs as the number of frames averaged increases, we calculated the radial power spectra of representative EBSPs from each map. The radial power spectrum offers a way to quantify the intensity and frequency of regular features in an image in any orientation, such as band edges (low frequency and long wavelength) and detector noise (high frequency and short wavelength) in EBSPs (Tokarski et al., 2021). Radial power spectra were calculated from two-dimensional fast fourier transforms of EBSPs from each map (Ruzanski, 2009). Specifically, we selected EBSPs from locations with the highest and lowest BC values in the map with one frame averaged that successfully indexed as olivine, then selected patterns from those same locations in the remaining five maps. The radial power spectra for EBSPs with high and low BC values are plotted versus wavelength in Figures 4a and 4b, respectively. As the number of frames averaged per pattern increases, the peak intensity of long-wavelength features in the EBSPs (i.e., band edges) increases while the intensity of short-wavelength features in the EBSPs (i.e., noise) decreases. The power spectra are broadly similar between the radial power spectra derived from the EBSPs with high BC and low BC values.

To quantitatively compare these spectra, we calculated the signal-to-noise ratio for each radial power spectrum, defined here as simply the ratio of the peak intensity at a wavelength of 70 pixels to the intensity at a wavelength of 3 pixels near the base of the peak. These signal to noise values are reported in Table 1. The signal-to-noise ratio increases with increasing number of frames averaged. There are no significant differences between the signal-to-noise ratios for spectra derived from EBSPs with high or low BC, although the signal-to-noise ratio for EBSPs with low BC appear to have a greater rate of improvement up to 15 frames averaged.

In Figures 4c and 4d, $m$ is plotted against the mean BC value for each map and against the signal-to-noise ratio of each radial power spectrum, respectively, both of which are related to the number of frames averaged during data collection. The value of $m$ decreases with increasing mean BC and with increasing signal-to-noise ratio of the power spectra. In fact, both of these relationships display similar trends as $m$ starts to asymptote with respect to each variable after the data points corresponding to maps with 10 frames averaged. The similarities between these trends suggest that both the mean BC and radial power spectra are good metrics by which to judge pattern quality. Importantly, because the mean BC is calculated and displayed during data collection by the AZtec software, it can be used as a reliable quantity to judge whether or not EBSPs will be of good enough quality to produce HR-EBSD data sets without feature-obscuring noise.

The trade off with using more frame averaging to achieve better pattern quality is the time for data collection. Because $m$ does not continue to change significantly in the maps with more than 10 frames averaged per point (Figures 4c and 4d) and few additional microstructural features in the maps are revealed by averaging more than 10 frames (Figure 2), 10 frames seems to be the optimal number of frames to average for the sample analyzed in the present study. This optimal number of frames may vary between materials and surface preparation techniques, so a better target for future analyses may instead be to aim for mean BC values greater than 100 or signal-to-noise ratios greater than 500 (Table 1). Most importantly, however, is that maps made for the purposes of comparison between samples have similar values of mean BC or signal-to-noise ratio, as large differences between these values in different maps can significantly impact other analyses, such as estimates of dislocation density from the restricted second moment.

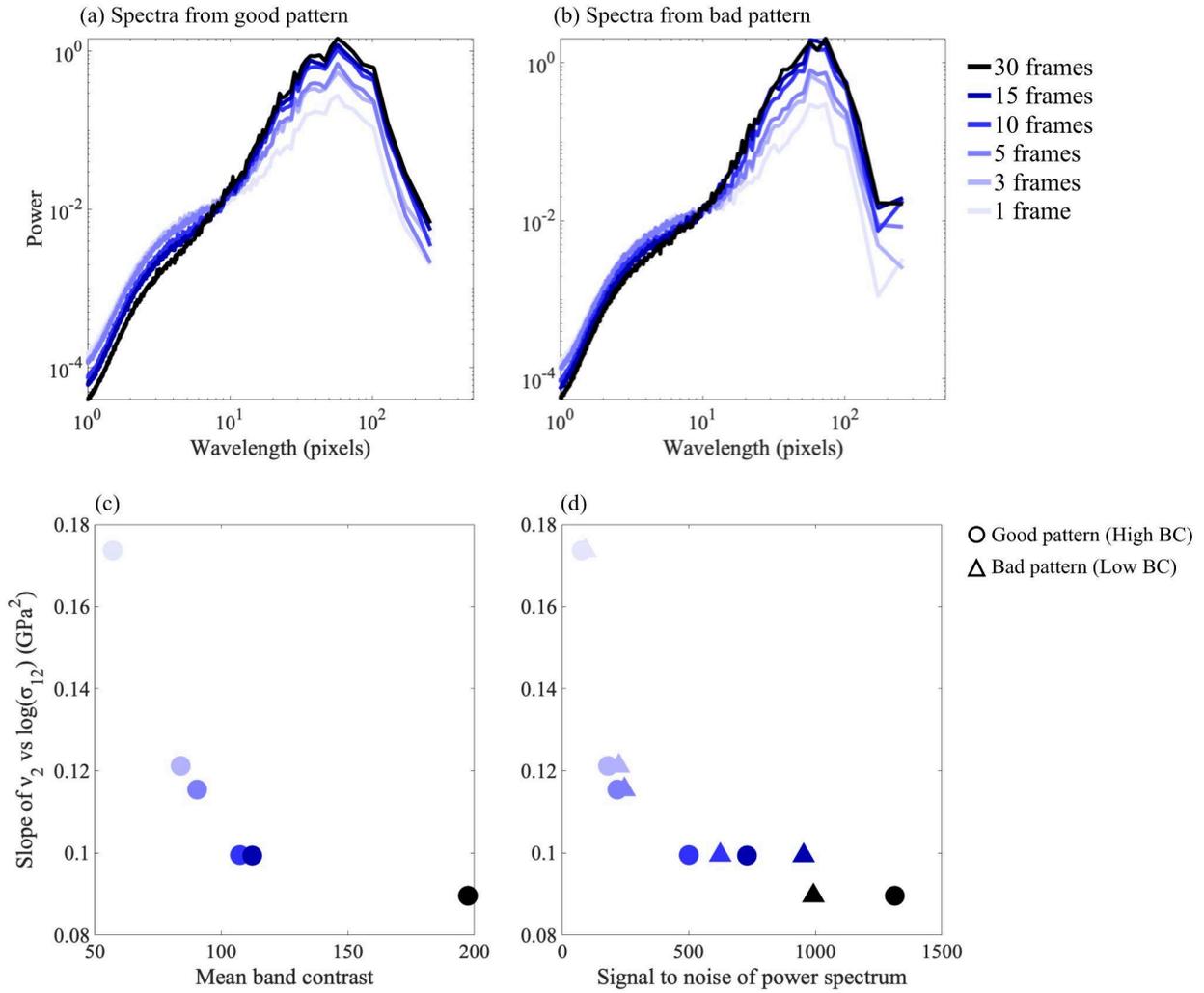

Figure 4. (a) and (b) Radial power spectra for EBSPs with high and low BC, respectively. Distributions are plotted on a log-log scale. (c) *m* versus mean BC and (d) *m* versus signal-to-noise ratio of the radial power spectra, respectively. In (d), data from the EBSP with high BC are plotted as circles, while those from the EBSP with low BC are plotted as triangles. In each panel, colors are the same as those in Figure 3.

**4.1.2 Pixel-by-Pixel Variation in HR-EBSD Maps**

To further examine how individual features in HR-EBSD maps change as the number of frames averaged increases, we calculated the difference between maps of intragranular stress heterogeneity and the difference between maps of GND density created with 1 frame and 15 frames averaged. This stress difference map is presented in Figure 5a and the GND difference map is presented in Figure 5b. We selected the data sets for 1 frame averaged and 15 frames averaged because they represent the largest difference in pattern quality. We chose not to compare to the maps created with 30 frames averaged because the mapped area did not overlap completely with the other maps. Due to the possibility of beam and/or sample drift during data collection, an image transformation was applied to the maps created with 15 frames averaged to align features, such as grain boundaries, in BC maps between the two sets of maps compared here. In Figures 5a and 5b, the greatest differences in stress heterogeneity and GND density are mostly along subgrain boundaries and in patches near grain boundaries that are associated with high stress magnitudes and high GND densities in Figure 2. The lattice in these regions is highly distorted, resulting in low BC compared to regions with less distortion (Figure 1).

To examine these changes more quantitatively, in Figures 5b and 5d we plot the differences in stress and GND density at each point versus the BC value at that same point in the map with 1 frame per pattern, respectively. These plots demonstrate that points with low BC had the greatest change in both stress heterogeneity and GND density with increasing frame averaging. For points with increasingly greater values of BC, the difference at each point is more varied but notably trends towards smaller differences. In other words, stress heterogeneity and GND density change the most at points with low BC, whereas points that already have high BC are more likely to stay the same with increasing frame averaging. As highly distorted regions, typically with low BC, are often the focus of HR-EBSD analyses, our results serve as an initial guide to employing frame averaging as a means to achieve the desired accuracy in HR-EBSD analyses.

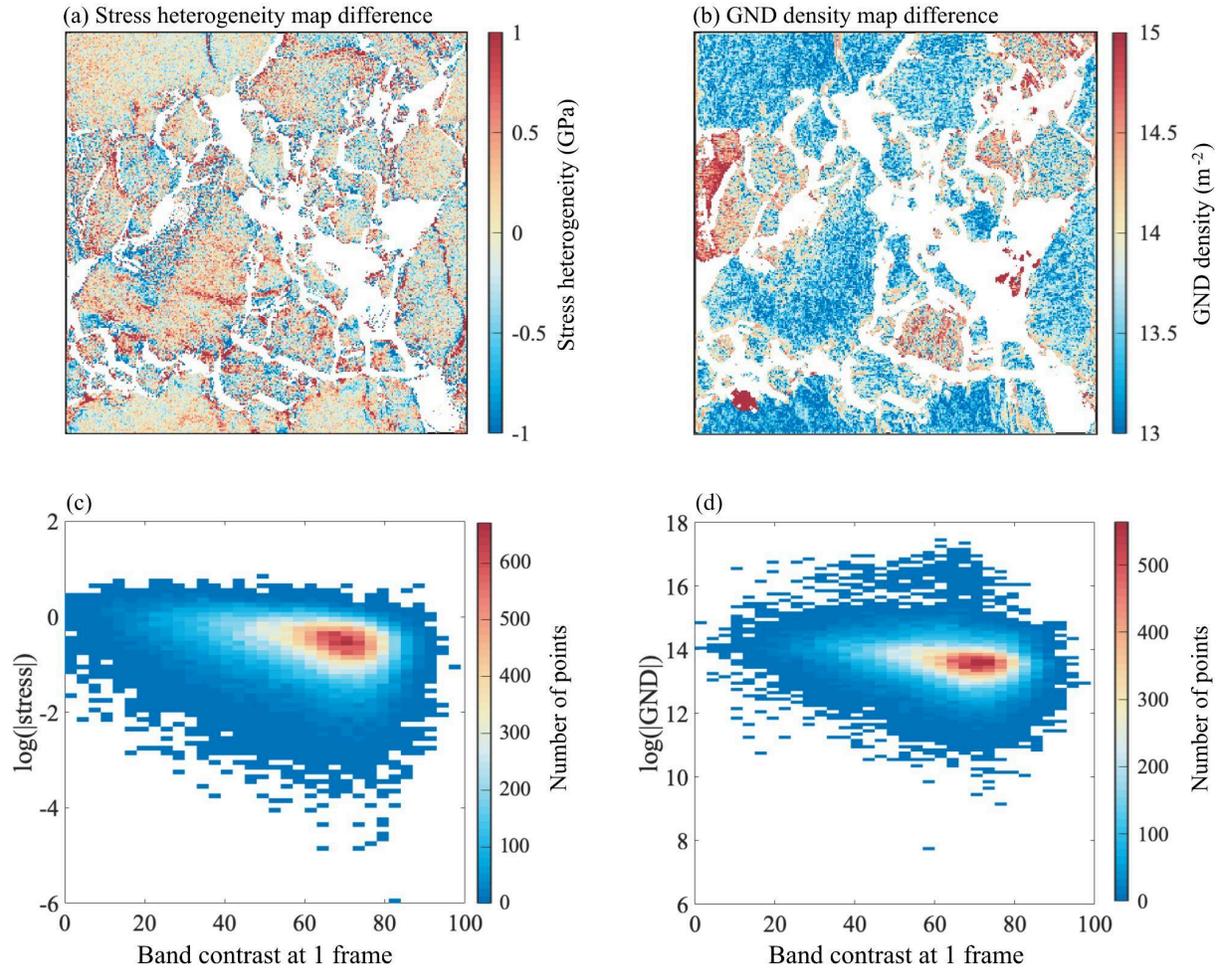

Figure 5. Difference in (a) stress heterogeneity and (b) GND density between maps with one frame averaged and 15 frames averaged. Two-dimensional histograms of (c) log(stress difference) and (d) log(GND difference) at each pixel in (a) and (b), respectively, versus the BC value at the corresponding pixel in the BC map with one frame averaged (Figure 1a). In (c) and (d) we took the absolute value of the stress difference and GND difference before plotting them on the histograms.

## 5. Conclusions

The pattern quality, or band contrast, of EBSPs affects the results of analysis with HR-EBSD. Specifically, analyses from maps with a higher mean band contrast reveal patches of stress heterogeneity and GND density obscured by noise in analyses from maps with lower mean band contrast among EBSPs. Employing frame averaging during data collection is an effective way to improve the quality of EBSPs and thereby reduce noise in HR-EBSD maps. Reducing noise in these regions is especially important, as dislocations of interest inherently distort the crystal lattice, reducing the band contrast. Although differences among surface preparation techniques (i.e., thickness of a carbon coat, quality of surface polish, etc.) will change the specifics of data collection, our analysis serves as a guide for improving the quality of analysis with HR-EBSD by targeting EBSD maps with mean band contrast values greater than 100 or EBSPs with signal-to-noise ratios greater than 500. Most important, however, is that maps made for the purposes of comparison between samples have similar values of mean band contrast or signal-to-noise ratio, as large differences between these values can alter the distribution of stresses or estimates of dislocation density from the stresses between different maps.

**Data Availability**

All data used in this study are available from GFZ data services (Wiesman et al., 2025)

**Acknowledgements**

The authors would like to thank Jessica Warren, Thomas Breithaupt, Charlie Gordon and Jessica White for helpful discussions throughout the data analysis. The research was supported by a UK Research and Innovation Future Leaders Fellowship grant number MR/V021788/1 awarded to DW. Microscopy was carried out at the Wolfson Electron Microscopy Suite at the University of Cambridge, which receives funding from the Cambridge Royce facilities grant EP/P024947/1 and Sir Henry Royce Institute recurrent grant EP/R00661X/1. Experiments were carried out at the Rock and Mineral Physics Laboratory at the University of Minnesota.